\documentstyle[aaspp4]{article}
\def\unsetyr{\def\oyear{\relax}\def\cyear{\relax}}
\def\setyr{\def\oyear{(}\def\cyear{)}}
\unsetyr
\def\jcite#1{\setyr\cite{#1}\unsetyr}
\def\rmmat#1{{\hbox{\rm #1}}}
\def\rmscr#1{\rmmat{\scriptsize #1}}
\newcommand{\be}{\begin{equation}}
\newcommand{\ee}{\end{equation}}
\newcommand{\ba}{\begin{eqnarray}}
\newcommand{\ea}{\end{eqnarray}}
\newcommand{\ie}{{\it i.e.~}}
\newcommand{\eg}{{\it e.g.~}}
\def\eqref#1{Equation~\ref{eq:#1}}
\def\figref#1{Figure~\ref{fig:#1}}

\begin{document}
\newcommand{\bfi}{{\bf B}} \newcommand{\efi}{{\bf E}}
\newcommand{\lel}{{\lambda_e^{\!\!\!\!-}}}
\newcommand{\me}{m_e}
\newcommand{\mcs}{{m_e c^2}}
\title{Powering Anomalous X-ray Pulsars by Neutron Star Cooling}
\author{Jeremy S. Heyl}
\authoremail{jsheyl@ucolick.org}
\author{Lars Hernquist\altaffilmark{1}}
\authoremail{lars@ucolick.org}
\affil{Lick Observatory,
University of California, Santa Cruz, California 95064, USA}
\altaffiltext{1}{Presidential Faculty Fellow}

\begin{abstract}
Using recently calculated analytic models for the thermal structure of
ultramagnetized neutron stars, we estimate the thermal fluxes from young
($t\sim 1000$ yr) ultramagnetized ($B \sim 10^{15}$ G) cooling neutron
stars.  We find that the pulsed X-ray emission from objects such as 1E
1841-045 and 1E 2259+586 as well as many soft-gamma repeaters can be
explained by photon cooling if the neutron star possesses a thin
insulating envelope of matter of low atomic weight
at densities $\rho < 10^{7}-10^{8}$ g/cm$^3$.  The total mass
of this insulating layer is $M \sim 10^{-11}-10^{-8} M_\odot$. 
\end{abstract}

\section{Introduction}

In recent years, several ``breaking'' (\cite{Mere95}) or ``anomalous''
(\cite{vanP95}) x-ray pulsars have been discovered (\cite{Vasi97b,Corb95}).
These objects typically have pulsed X-ray emission with steadily
increasing periods $\sim$ 10 s, X-ray luminosities $\sim
10^{35}-10^{36}$ erg/s, soft spectra, and no detected companions or
accretion disks.  Furthermore, they are typically observed through hydrogen
column densities $\sim 10^{22}$ cm$^{-2}$ indicating that they are not
common.   \jcite{Vasi97b} describe several of these sources and present
observations of 1E 1841-045, whose properties are characteristic of
this class of objects.

Specifically, \jcite{Vasi97b} use archival observations of the
supernova remnant Kes 73 obtained with the ASCA and ROSAT satellites.  They
find that the x-ray source in the center of the SNR, 1E 1841-045,
had a period of 11.766684 s from the ASCA data taken in 1993 October.
The ROSAT data of 1992 March is best fitted with a period of 11.7645 s,
yielding a period derivative of ${\dot P} \simeq 4.73 \times 10^{-11}
\rmmat{s s}^{-1}$ and a characteristic spin-down age of 4,000 yr --
close to the estimated age for Kes 73 of 2,000 yr.  Using these values and
assuming that magnetic dipole radiation dominates the spin-down,
they estimate the dipolar field strength of the neutron star to be
$\sim 10^{15}$ G, well above the quantum critical field, $B_{cr}
\approx 4.4\times 10^{13}$ G.  Other anomalous x-ray pulsars (AXPs)
generally have small ages and long periods, leading one to derive similar
field strengths. 

With x-ray luminosities $L_X \sim 10^{35}-10^{36}$ erg/s, AXPs are
underluminous relative to accretion powered X-ray pulsars and are
generally isolated.  For 1E 1841-045, \jcite{Vasi97b} estimate a
spin-down power of $10^{33}$ erg/s which falls short of the observed
luminosity.  They also argue that although 1E 1841-045 has a period
near the equilibrium spin period for a young pulsar with $B \sim
10^{12}$ G and $L_X \sim 10^{35}$ erg/s, only an unlikely evolutionary
process could spin down the neutron star to this rate within the 2,000
year age of Kes 73.  They suggest that 1E 1841-045 may be powered by
magnetic field decay in a dipolar field of strength $B \sim 10^{15}$ G
(\cite{Thom96,Gold92}).

In this {\it Letter}, we propose a natural explanation for the
observed X-ray emission from AXPs.  Neutron stars with ages $\sim$
1,000 yr and magnetic fields $B\gtrsim 10^{15}$ G have thermal
emission in the X-ray-band with total luminosities $\sim 10^{35}$
erg/s, if their surface layers consist of light-weight material, such
as hydrogen and helium.  In previous papers, we have developed an
analytic model for ultramagnetized neutron star envelopes
(\cite{Heyl97a}) and calculated the emission through iron envelopes
and showed how a strong magnetic field affects neutron-star cooling
(\cite{Heyl97b}).  Here, we will examine the properties of
ultramagnetized hydrogen and helium envelopes and draw parallels with
the observed properties of AXPs.

\section{Model Envelopes}

\jcite{Heyl97a} have developed analytic models for ultramagnetized
neutron star envelopes and find that the flux transmitted through the
envelope is simply related to the direction and strength of the
magnetic field and to the core temperature ($T_c$).  Specifically,
these models apply only below densities $\rho_\rmscr{max}$ at which
the ground Landau level becomes filled; \ie $2.2 \times 10^8 - 7.1
\times 10^9$ g/cm$^3$ for polar fields, $B=10^{15} - 10^{16}$ G.

Here we consider hydrogen and helium envelopes with $B\sim 10^{15}$ G
and core temperatures expected for cooling by the modified URCA
process after $\sim$ 1,000 yr.  Although our models do not extend to
the high densities traditionally associated with the isothermal core
of a neutron star ($\rho \gtrsim 10^{10}$ g/cm$^3$), for $B\sim
10^{15}$ G and $T_c \sim 10^8-10^{8.5}$ K, the envelope is nearly
isothermal at $\rho_\rmscr{max}$.  We denote the temperature at this
density by $T_\rmscr{max}$ and for this analysis take it to be equal
to the core temperature.

The envelope models are calculated using a plane-parallel, Newtonian
approximation.  In this approach, the core temperature is a
function of $F/g_s$, $B$ and $\psi$ (the angle between the radial and
field directions).  $F$ is the transmitted heat flux, $g_s$ is the
surface gravity, and all of these values are taken to be in the frame of
the neutron star surface.  For such strong fields, the models have a
simple dependence on the angle $\psi$, \ie $F/g_s \propto \cos^2\psi$
(\cite{Heyl97a,Shib95,Shib96}) and furthermore the flux for a fixed core
temperature is approximately proportional to $B^{0.4}$.  With these two
facts, we find that the average flux over the surface of a neutron star
with a dipole field configuration is 0.4765 times its peak value at the
magnetic poles.

For a given core temperature, magnetized envelopes transmit more heat
than unmagnetized ones (\eg \cite{Hern85}).  Furthermore, as
\jcite{Hern84b} found, the relationship between the core temperature
and the transmitted flux is strongly sensitive to the composition of
the degenerate portion of the envelope, specifically iron 
insulates the core much more effectively that hydrogen or helium.
Because of the intense magnetic field, the luminosities for a given
core temperature are even larger than those found by \jcite{Pote97}.

\section{Luminosity Evolution}

We have calculated the expected luminosity of an ultramagnetized neutron
star as a function of time.  We assume that the core of the neutron star
cools by only the modified URCA process (\eg \cite{Shap83}),
\be
\Delta t (\rmmat{URCA}) \simeq 1 
\left ( \frac{\rho}{\rho_\rmscr{nuc}} \right )^{-1/3} T^{-6}_{c,9}(f)
\left \{ 1 - \left [ \frac{T_{c,9}(f)}{T_{c,9}(i)} \right ]^6 \right \}
\rmmat{yr} ,
\ee
where $T_x=T/10^x$ K and $\Delta t$ is an estimate of the age of the
neutron star.  For simplicity, we will assume that the initial
temperature is much larger than the current temperature $T_{c}(f)$.

If one ignores gravitational redshift effects which depend on the
radius of the neutron star, the photon luminosity is given by
\be
L_\gamma = 4 \pi G M \frac{F}{g_s} \approx 9.5 \times 10^{32} 
\frac{{\bar T}_{\rmscr{eff},6}^4}{g_{s,14}} \frac{M}{M_\odot} \rmmat{erg
s}^{-1} 
\ee
We will take $M=1.4 M_\odot$.  

\begin{figure} 
\plotone{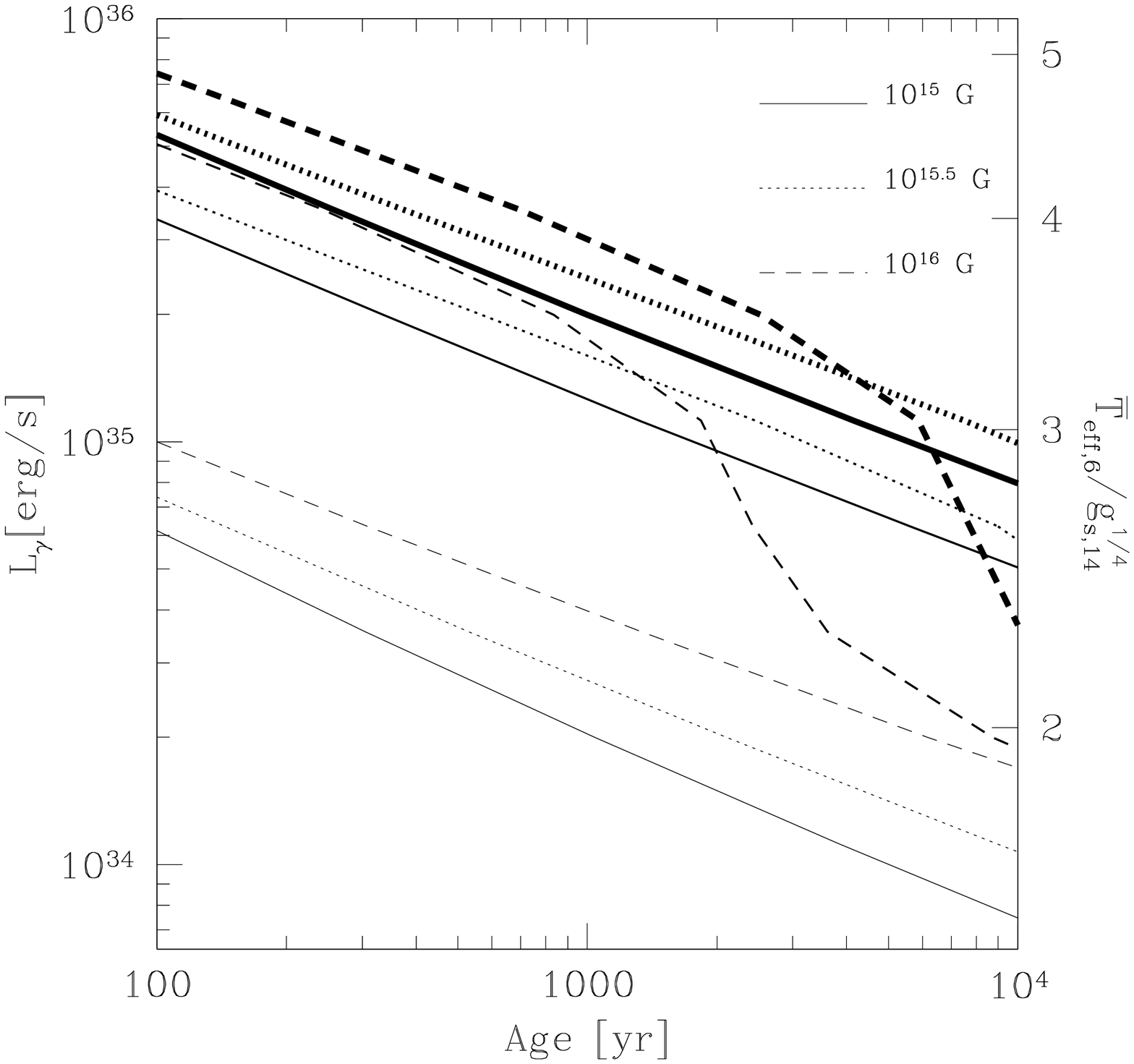}
\caption{
Photon luminosity and mean effective temperature as a function of the
age of the neutron star.  The upper bold curves trace the evolution for
hydrogen envelopes, the intermediate curves give the evolution for
helium envelopes and the lowest light curves follow the cooling through
iron envelopes.   The values of $\rho_\rmscr{max}$ for hydrogen
envelopes are given in the legend.
}
\label{fig:kesfig}
\end{figure}

\figref{kesfig} depicts the photon luminosity and mean effective
temperature as a function of time for several field strengths with iron,
helium, and hydrogen envelopes.  We see that the photon luminosity increases
with increasing magnetic field, but the composition of the envelope is 
a more important effect.  The luminosity through a hydrogen envelope is ten
times larger than through iron and 1.6 times larger than through
helium.  The relationship between core temperature and flux is most
sensitive to the thermal conductivity in the liquid portion of the
degenerate envelope (\cite{Gudm82}).  In this zone, the conductivity
is proportional to $Z^{-1} $; consequently, low $Z$ envelopes conduct
heat more readily.

To lowest order, for a given magnetic field strength and core
temperature, the transmitted flux is proportional to $Z^{-2/3}$.  The
dependence on $Z/A$ is weaker.  The flux dips sharply 
for hydrogen and helium with $B=10^{16}$ G, when the material near the
degenerate-non-degenerate interface begins to solidify.  At a
given density the conductivity due to degenerate electrons is
smaller in the solid state than in the liquid phase.

We note that our assumption that the magnetic field has no explicit
influence on the phase of the matter is problematic for these intense
fields.  For example, the magnetic field may well alter the state of
the material in the solid portion of the crust, particularly when the
magnetic stress exceeds the yield stress of the lattice.  The shear
modulus of the lattice is $\mu \sim (Z e)^2 n_Z^{4/3}$, where $n_Z$ is
the ion number density (\cite{Rude72}).  Assuming that the lattice
breaks when the strain angle $\chi_\rmscr{strain} \sim 10^{-2}$,
magnetic stresses dominate when (\eg \cite{Blan82}; \cite{Blan83})
\be
B \gtrsim 1.2 \times 10^6 Z A^{-2/3} \left( \frac{\rho}{1
\rmmat{~g cm}^{-3}} \right)^{2/3} \rmmat{G}, 
\ee
which is satisfied throughout the crust for fields $B \gtrsim 10^{14}$
G.

Even for the weakest field considered, $B=10^{15}$ G, the photon
luminosity through a hydrogen envelope is $1.5 \times 10^{35}$ erg/s for
an age of 2,000 yr.  This is comparable to both the pulsed X-ray pulsed
luminosity of 1E 1841-045 of $L_X \approx 5 \times 10^{34} d_7^2$ erg/s
($d_7$ is the distance to the source divided by 7 kpc) and the total
X-ray luminosity of $3.5 \times 10^{35} d_7^2$ erg/s, given
observational uncertainty.

The photon luminosity is nearly large enough to account for the
luminosity in the quiescent state of the soft gamma repeater (SGR
0526-66).  Several authors have advanced the view that strongly magnetized
neutron star power SGRs.  For example, \jcite{Thom95} propose that
soft gamma repeaters (SGRs) are powered by magnetic reconnection
events near the surfaces of ultramagnetized neutron stars.
\jcite{Ulme94} finds that a strong magnetic field can
explain the super-Eddington radiation transfer in SGRs. \jcite{Roth94}
estimate the luminosity of SGR 0526-66 in the quiescent state to be
approximately $7\times 10^{35}$ erg/sec.  Since SGR 0526-66 is located
in a supernova remnant, they can also estimate the age of the source
to be approximately 5,000 years.  For this age and a magnetic field of
$B=10^{15.5} G$, we find a photon luminosity of $1.5\times 10^{35}$
erg/sec, not far short of the SGR's quiescent luminosity.

\section{Discussion}

To determine the viability of our model for AXP emission, we must
ascertain whether or not neutron stars with $B\sim 10^{15}$ G are likely
to have an insulating envelope of the required mass in low atomic weight
matter.  From numerical experimentation, we have found that the
relationship between the core temperature and the effective temperature
is most sensitive to the composition in the degenerate, liquid portion
of the envelope; \ie the ``sensitivity'' region found by \jcite{Gudm82}.
The density in this zone is approximately in the range $10^6 - 10^7$
g/cm$^3$. 

We can appeal to the equation of hydrostatic equilibrium to calculate
the total mass up to a certain density.  If we assume 
that the envelope is thin, we have
\be
\Delta M = \frac{4 \pi R^2}{g_s} P,
\ee
where $R$ is the radius of the neutron star and $P$ is the pressure at
the given depth.  In the region of interest, the pressure is supplied
by degenerate electrons occupying the ground Landau level.  In a
strong magnetic field, electrons become relativistic at
\be
\rho \approx 3.3 \times 10^7 B_{15} \mu_e \rmmat{g/cm}^{3},
\ee
where $B_{15}=B/10^{15}$ G  and  $\mu_e=A/Z$ is
the mean atomic weight per electron (\eg \cite{Heyl97a}).

For nonrelativistic electrons the equation of state is
\be
P_e = 1.5 \times 10^{20} B_{15}^{-2} \mu_e^{-3}
\rho_6^3  \rmmat{dynes/cm}^2 \rmmat{~for~} \rho_6 \ll 33 B_{15} \mu_e,
\ee
where $\rho_6=\rho/10^6$ g/cm$^3$.

For densities $\rho_6 \gg 33 B_{15} \mu_e$, the electrons are
relativistic, and we obtain the equation of state,
\be
P_e = 7.5 \times 10^{21} B_{15}^{-1} \mu_e^{-2}
\rho_6^{2} \rmmat{dynes/cm}^2 .
\ee
Therefore, depending on the maximum density of the insulating
layer, we obtain for its total mass,
\be
\Delta M = 7.1 \times 10^{-15} R_6^4 
\left ( \frac{M}{M_\odot} \right )^{-1} B_{15}^{-2} \mu_e^{-3} \rho_6^3
 M_\odot 
\ee
in the non-relativistic limit where $M$ is the mass of the neutron star and
$R_6 = R/10^6$ cm.  In the relativistic limit, we obtain
\be
\Delta M = 3.5 \times 10^{-13} R_6^4 
\left ( \frac{M}{M_\odot} \right )^{-1} B_{15}^{-1} \mu_e^{-2} \rho_6^2
M_\odot .
\ee
At a minimum, the low-$Z$ insulating layer must extend into the
degenerate portion of the envelope where $\rho_6 \sim 10$.  The
non-relativistic estimate for the mass of this layer yields
$\sim 10^{-11} M_\odot$ for $B=10^{15}$ G.  If the layer
extends to the density at which the first Landau level fills ($\rho_6
\sim 200$) well into the relativistic regime, $\Delta M \sim 10^{-8}
M_\odot $ for $B=10^{15}$ G.

The high temperatures accompanying core collapse are generally thought
to process the material that will comprise the neutron star envelope
to nuclear statistical equilibrium; therefore, the envelope is
expected to consist of $^{56}$Fe and heavier nuclei at higher
densities (\eg \cite{Shap83}).  A hydrogen or helium envelope must contain
material accreted by the neutron star from the interstellar medium or
from the ``detritus'' of the supernova.  Even with an intense magnetic
field to dramatically increase the Alfven radius, a neutron star
would require $\sim 10^6$ yr to accrete $\sim 10^{-11} M_\odot$ from the
ISM, unless the neutron star were traveling through an unusually
dense region.  For example,
\be
{\dot M} \sim 10^{-14} 
    n_4^{7/9} R_6^{4/3} B_{15}^{4/9} v_7^{7/9}
    \left ( \frac{M}{M_\odot} \right )^{2/3}
    M_\odot/\rmmat{yr},
\ee
where $n_4$ is the number density of the ISM in units of $10^4
\rmmat{cm}^{-3}$, $v_7=v/10^7$ cm/s, and $v$ is the velocity of the
neutron star.  In such a dense medium, the neutron star could accrete
the minimum required $10^{-11} M_\odot$ within $\approx$ 1,000 years.
If the nascent neutron star received a kick during the supernova
explosion, it could accrete material from the high density remnant of
the progenitor star.  \jcite{Dunc92} argue that a strong magnetic
field $B\sim 10^{15}$ G can cause the neutrino emission from the
nascent neutron star to be anisotropic; consequently, the neutron star
would recoil at $v \sim 100$ km/s relative to center of the 
explosion.

\jcite{Chev89} modeled the accretion by a nascent neutron star or
black hole from the exploded envelope of SN 1987A.  He found that
$\sim 0.1 M_\odot$ will fall back within several days of the
explosion.  For a typical type II supernova which has an extended
envelope, the total accreted mass may be a factor of 100 smaller,
yielding a total of $10^{-3} M_\odot$ of material to comprise the
insulating layer.  Although much of this material will have been
processed during the evolution of the star, the X-ray and gamma-ray
light curves of SN 1987A as well as numerical simulations indicate
that heavy and light elements mix during the explosion (\cite{McCr93}
and references therein); any unprocessed remnants of the progenitor
star's hydrogen envelope would quickly float to the surface
(\cite{Lai97}) and form a hydrogen envelope on the neutron star.  Even
in an intense magnetic field, pycnonuclear fusion reactions do not
proceed quickly enough to substantially process the accreted material
within 1,000 yr (\cite{Heyl96a}; \cite{Lai96}).  If insufficient
hydrogen fell back, the photon luminosity for a helium envelope is
only one third lower than that of a hydrogen envelope; consequently,
even a Type Ib supernova could result in a cooling-powered neutron
star with sufficient luminosity to be observed as an AXP.

\jcite{Zavl96} calculate several model atmospheres for cooling neutron
stars including hydrogen and helium for weak magnetic fields ($B \sim
10^8 - 10^{10}$ G), and compare the emergent spectra with a blackbody
distribution.  Because the opacity of a hydrogen or helium atmosphere
drops quickly with photon energy, high energy photons originate from
deeper and hotter layers of the atmosphere; consequently, hydrogen and
helium atmospheres have spectra which peak at a higher frequency and
have stronger Wien tails than a blackbody spectrum with the same
effective temperature (\cite{Roma87}).  \jcite{Pavl96} convolve the
spectra from neutron star atmospheres consisting of hydrogen with the
ROSAT response function and find that the fitted values of
$T_\rmscr{eff}$ are 4.17 times higher than the model value.  The
observed best-fit blackbody temperature for the spectrum of
1E~2259+586 is 0.45 keV.  This is about twice the effective
temperature of our model envelopes.  The processing of the radiation
by the hydrogen atmosphere may account for this shift.  

Furthermore, in the energy range of 1--10 keV, hydrogen spectra tend to be
less steep ($S_\nu \propto \nu^{-5}$) than the equivalent blackbody
spectra.  Although they are not as shallow as the non-thermal tail observed
in 1E~2259+586, $S_\nu \propto \nu^{-3}$, mechanisms in the magnetosphere
probably contribute to the high-energy emission.

An intense magnetic field induces a temperature variation across the
surface of the neutron star (\cite{Heyl97a}).  This gradient combined
with the limb darkening manifest in models of magnetized neutron
star atmospheres (\cite{Roma87}, \cite{Zavl96}) can naturally explain
the observed pulsed fraction of Kes 73 of 35\% (\cite{Vasi97b}).

\section{Conclusions}

We find that young ($t\sim 1,000$ yr) neutron stars with strong
magnetic fields ($B \sim 10^{15}$ G) and hydrogen or possibly helium
envelopes have photon luminosities similar to those observed from
anomalous X-ray pulsars and soft-gamma repeaters in their quiescent
state.  The total mass of the insulating layer is $10^{-11} - 10^{-8}
M_\odot$.  A strongly magnetized neutron star could accrete enough
material from the ISM within 1,000 years only if the ISM is sufficiently
dense, $n \gtrsim 10^4 \rmmat{cm}^{-3}$.  However, sufficient material
is expected to fall back onto the neutron star surface following the
explosion of massive stars.

\acknowledgements

The work was supported in part by a National Science Foundation
Graduate Research Fellowship and the NSF Presidential Faculty Fellows
program. 

\bibliography{ns}
\bibliographystyle{jer}

\end{document}